\documentclass[cameraready]{Interspeech}

\title{RT-SEMamba: Real-Time Speech Enhancement Mamba via
Progressive Knowledge Distillation}

\usepackage[utf8]{inputenc}
\usepackage{times}
\usepackage{graphicx}
\usepackage{booktabs}
\usepackage{amsmath}
\usepackage{tikz}
\usepackage{hyperref}
\usepackage{xcolor}  

\author[affiliation={1,2}]{Rong}{Chao}
\author[affiliation={5}]{Sung-Feng}{Huang}
\author[affiliation={3}]{Moreno}{La Quatra}
\author[affiliation={4}]{Sabato Marco}{Siniscalchi} 
\author[affiliation={2}]{\\ Wen-Huang}{Cheng}
\author[affiliation={5}]{Szu-Wei}{Fu}
\author[affiliation={1}]{Yu}{Tsao}

\address{
$^{1}$Academia Sinica, Taiwan\ \ 
$^{2}$National Taiwan University, Taiwan \\
$^{3}$Kore University of Enna, Italy\ \ 
$^{4}$University of Palermo, Italy \ \ 
$^{5}$NVIDIA
}

\email{roychao@cmlab.csie.ntu.edu.tw, yu.tsao@citi.sinica.edu.tw}


\usepackage{comment}
\usepackage{amsmath,amssymb}

\begin{document}

\maketitle






\begin{abstract}

We present RT-SEMamba, a fully causal speech enhancement (SE) model built upon causal time--frequency Mamba blocks. Unlike Transformer-based architectures that rely on a growing key--value cache, Mamba propagates a fixed-size recurrent state per layer, enabling memory- and bandwidth-efficient long-form inference. We further introduce a progressive knowledge distillation (KD) strategy that compresses an 8-layer teacher into a shallow 1-layer student by jointly distilling complex spectral outputs and intermediate representations. On Voicebank-DEMAND, the 8-layer RT-SEMamba achieves 3.32 PESQ with a 25\,ms algorithmic latency constraint, and the distilled 1-layer student improves over a naive 1-layer baseline from 3.06 to 3.18 PESQ while preserving the same steady-state RTF, delivering a 2.64$\times$ speedup over the teacher. These results demonstrate that state-space models with progressive KD provide a competitive quality--latency trade-off for real-time SE.

\end{abstract}

\section{Introduction}
\label{sec:intro}

Real-time speech enhancement (SE) is a key component in interactive audio applications such as hearing aids~\cite{HA,ahmed2025neuroamp}, cochlear implants~\cite{lai2016deep}, AR/VR communication~\cite{yang2022audio}, and live teleconferencing~\cite{hsu2022learning}. In these scenarios, the SE front-end must satisfy tight algorithmic latency and real-time factor (RTF) constraints while robustly handling non-stationary noise and reverberation. Unlike non-causal SE, streaming SE can only rely on past and current observations, yet still needs to deliver artifact-free speech under limited compute and memory resources~\cite{realtimeSE,schroter2022deepfilternet}.

Modern SE systems are typically formulated as supervised regression from noisy waveforms or time--frequency representations to their clean targets~\cite{SE, o2024speech, wang2018supervised, kolbaek2020loss}. Convolutional, recurrent, and hybrid convolutional–recurrent networks~\cite{FCN,LSTM,li2020speech,tan2018convolutional}, as well as Transformer-style architectures~\cite{Trans,CMGAN, mp_senet}, achieve strong performance in non-causal benchmarks. Phase-aware, complex-spectrum, sub-band, and metric-oriented models further improve time--frequency enhancement quality~\cite{yin2020phasen,hu2020dccrn,hao2021fullsubnet,fu2019metricgan,fu2021metricgan+}. More recently, diffusion-based generative models have been explored as an alternative SE formulation~\cite{diffSE2022, Richter2023, scheibler2024universal, welker22_interspeech, li2024diffusion, gonzalez2024investigating}. However, many Transformer- and diffusion-based SE systems are evaluated with non-causal future context, making low-latency streaming deployment less straightforward.



To better match streaming requirements, several works have proposed explicitly causal and low-latency SE architectures~\cite{realtimeSE, streamingSE, schroter2022deepfilternet}. Waveform-domain models demonstrate real-time operation on commodity hardware~\cite{realtimeSE}, and causal front-ends built on self-supervised representations show that semantic priors can partially compensate for limited context~\cite{streamingSE}. 



More recently, Mamba-style selective state-space models~\cite{gu2023mamba} have been introduced as linear-complexity alternatives to self-attention, and have been adapted to time--frequency speech enhancement and separation~\cite{chao2024investigation,li2024spmamba,wang2025mamba, avenstrup2025sepmamba, kim2025mamba, chao2025universal}. 
By propagating a fixed-size recurrent state per layer, Mamba naturally supports long-form sequence modeling with constant memory and bandwidth; unlike Transformer/Conformer encoders that maintain a growing key--value cache, it realizes streaming as a pure recurrent update that avoids cache growth and reduces memory traffic on edge devices. These Mamba-based systems are primarily evaluated in non-causal settings and do not explicitly enforce streaming constraints.

In this work, we revisit SEMamba~\cite{chao2024investigation} from the perspective of strict streaming deployment and propose a real-time variant, RT-SEMamba. Unlike the original SEMamba, which contains non-causal components and is evaluated offline, RT-SEMamba is made fully causal and operates in an online streaming regime, making it directly applicable to low-latency SE. We further introduce a depth-compression scheme~\cite{hinton2015distilling} in which a deeper 8-layer causal Time-Frequency Mamba (cTF-Mamba) teacher distills its capability into a lightweight 1-layer student. On the VoiceBank-DEMAND (VCTK-DEMAND) dataset~\cite{valentini2016investigating}, the distilled RT-SEMamba (8-layer$\rightarrow$1-layer) consistently improves over a naive 1-layer baseline on common objective and perceptual metrics, while substantially narrowing the gap to the 8-layer teacher. At the same time, it matches the real-time factor of the 1-layer baseline and reduces steady-state RTF by more than 60\% compared with the 8-layer teacher under the same algorithmic latency constraint of 25 \, ms, indicating that RT-SEMamba is suitable for streaming deployment.



\begin{figure*}[htb]
    \centering
    \centerline{\includegraphics[width=16.3cm]{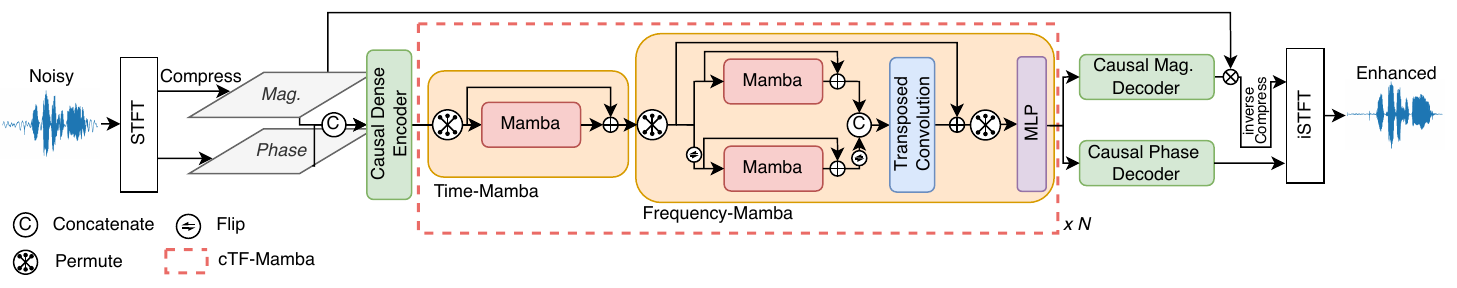}}
    \caption{Architecture of the proposed RT-SEMamba.}
    \label{fig:architecture}
    \vspace{-4mm}
\end{figure*}


\section{Related Work}
\label{sec:related_work}

\subsection{Mamba and SSM-based sequence models}

Mamba~\cite{gu2023mamba} is a selective state space sequence model that replaces self-attention with a lightweight linear recurrence. In its simplest form, an SSM maps an input sequence $\mathbf{x} = \{x_n\}_{n=1}^N$ to an output sequence $\mathbf{y} = \{y_n\}_{n=1}^N$ through a latent state $\mathbf{h}$:
\begin{equation}
    \mathbf{h}_n = \bar{\mathbf{A}}\,\mathbf{h}_{n-1} + \bar{\mathbf{B}}\,x_n,
    \qquad
    y_n = \mathbf{C}\,\mathbf{h}_n,
\end{equation}
where $\bar{\mathbf{A}}$ and $\bar{\mathbf{B}}$ are discretized state transition matrices and $\mathbf{C}$ is an output projection. This structure admits both parallel convolutional and recurrent implementations, yielding linear time and memory complexity for long sequences.


Building on this backbone, Mamba-style blocks have been applied to time--frequency speech enhancement and separation~\cite{chao2024investigation,li2024spmamba,wang2025mamba, chao2025universal}. In contrast, RT-SEMamba targets causal streaming speech enhancement with explicit state propagation and progressive distillation from a deeper causal teacher.

\section{RT-SEMamba Architecture}
\label{sec:method}

We operate in the complex STFT domain. For a noisy waveform $x(t)$, we compute
\begin{equation}
\mathbf{X}(f,\tau) = \mathrm{STFT}(x(t)) = \mathbf{M}(f,\tau)\,e^{j\mathbf{P}(f,\tau)},
\end{equation}
where $\mathbf{M}$ and $\mathbf{P}$ denote magnitude and phase. Both teacher and student networks take $(\mathbf{M},\mathbf{P})$ as input and predict $(\hat{\mathbf{M}}, \hat{\mathbf{P}}, \hat{\mathbf{C}})$, where $\hat{\mathbf{C}}=\hat{\mathbf{M}}e^{j\hat{\mathbf{P}}}$ is the complex reconstructed spectrum. The enhanced waveform is obtained via inverse STFT from $(\hat{\mathbf{M}}, \hat{\mathbf{P}})$.

\subsection{Causal Formulation and Real-Time Inference}
\label{sec:causal_setup}

RT-SEMamba is implemented in a fully causal manner, with several modifications to the non-causal SEMamba baseline~\cite{chao2024investigation}; its overall architecture is shown in Fig.~\ref{fig:architecture}. The front-end uses causal STFT/iSTFT at 16\,kHz with window size $W=400$, hop $H=100$, and \texttt{center=False} for both analysis and synthesis, so the algorithmic latency is bounded by one window (25\,ms), and no extra spectral normalization is applied. All temporal convolutions in the encoder and decoders use asymmetric causal padding: for kernel size $K$ we pad $(K-1)$ zeros on the left and none on the right, with missing past context at the beginning of an utterance filled by zeros. We also replace \texttt{InstanceNorm2d} with channel-wise LayerNorm with causal padding along the time axis, and insert an additional MLP after each cTF-Mamba block to improve per-frame modeling capacity. Time Mamba itself is made uni-directional along time: frames are processed sequentially over $t=1,\dots,T$ without lookahead, while sequence modeling along the frequency axis within each frame remains bidirectional. In our experiments, we deploy the model in an online streaming configuration, described next.

\subsubsection{Streaming inference setup}
For streaming inference, the model operates online in a 1-frame-in/1-frame-out mode. To avoid recomputing past frames, we propagate a small set of states across time: a temporal frame buffer storing the previous $(K-1)$ feature frames required by causal temporal convolutions in the encoder and decoders, a \texttt{conv\_state} buffer that keeps the internal buffer of the depthwise 1D convolution preceding the SSM over the last $(d_{\text{conv}}-1)$ time steps, and an \texttt{ssm\_state} that holds the recurrent hidden state $h_t$ of the selective state-space model. At each time step $t$, the model consumes one input frame, updates these states, and emits one enhanced frame, discarding obsolete values. In this setup, the model runs without lookahead, avoids redundant recomputation of past frames, and keeps per-frame compute and memory effectively independent of the sequence length $T$.

\begin{table}[t]
\centering
\caption{Model size, computational complexity per second (MACs (G/s)), and steady-state RTF versus number of cTF-Mamba layers. RTF is measured on a single NVIDIA RTX 5090 GPU, after warm-up.}
\label{tab:rtf_layers}
\begin{tabular}{c c c c}
\toprule
\# cTF-Mamba Layers & Params (M) & MACs (G/s) & RTF \\
\midrule
1-layer & 1.05 & 20.56 & 0.11 \\
2-layer & 1.29 & 24.39 & 0.13 \\
3-layer & 1.53 & 28.21 & 0.16 \\
4-layer & 1.77 & 32.04 & 0.19 \\
5-layer & 2.01 & 35.87 & 0.22 \\
6-layer & 2.25 & 39.70 & 0.24 \\
7-layer & 2.50 & 43.52 & 0.27 \\
8-layer & 2.74 & 47.35 & 0.29 \\
\bottomrule
\vspace{-8mm}
\end{tabular}
\end{table}


\subsection{Knowledge Distillation}

We use an 8-layer RT-SEMamba as the teacher $\mathcal{T}$ and a 1-layer RT-SEMamba as the student $\mathcal{S}$. For a noisy input $(\mathbf{M},\mathbf{P})$,
\begin{align}
(\hat{\mathbf{M}}^{t}, \hat{\mathbf{P}}^{t}, \hat{\mathbf{C}}^{t})
&= \mathcal{T}(\mathbf{M},\mathbf{P}), \\
(\hat{\mathbf{M}}^{s}, \hat{\mathbf{P}}^{s}, \hat{\mathbf{C}}^{s})
&= \mathcal{S}(\mathbf{M},\mathbf{P}).
\end{align}
The teacher parameters are frozen. To accelerate convergence, we initialize the student's encoder, decoder, and single TF-Mamba block with the corresponding weights from the pre-trained teacher.

\subsubsection{Output-level distillation}

We align student predictions with teacher outputs at magnitude, phase, and complex levels:
\begin{align}
\mathcal{L}_{\mathrm{KD}}^{\mathrm{out}}
&=
w_{\mathrm{mag}}
\lVert
\hat{\mathbf{M}}^{s}
-
\hat{\mathbf{M}}^{t}
\rVert_2^2
+
w_{\mathrm{pha}}
\lVert
\hat{\mathbf{P}}^{s}
-
\hat{\mathbf{P}}^{t}
\rVert_2^2 \\
&\quad+
w_{\mathrm{com}}
\lVert
\hat{\mathbf{C}}^{s}
-
\hat{\mathbf{C}}^{t}
\rVert_2^2.
\end{align}
We set
$w_{\mathrm{mag}}=1.0$,
$w_{\mathrm{pha}}=0.3$,
$w_{\mathrm{com}}=0.5$.

\subsubsection{Intermediate feature distillation}

Let $\mathbf{H}_i^{t}$ be the output of the $i$-th cTF-Mamba block in the teacher:
\begin{equation}
\mathbf{H}_i^{t}
=
\Phi_i^{t}(\mathbf{Z}_{i-1}^{t}),
\quad i=1,\dots,8,
\end{equation}
and let $\mathbf{H}^{s}$ denote the student block output. Since the student has only one block, we aggregate teacher features as
\begin{equation}
\mathbf{H}_{\mathrm{agg}}^{t}
=
\frac{1}{8}
\sum_{i=1}^{8}
\mathbf{H}_i^{t}.
\end{equation}
To reduce scale mismatch, we normalize features per sample:
\begin{equation}
\mathrm{Norm}(\mathbf{X})
=
\frac{\mathbf{X}-\mu(\mathbf{X})}
{\sigma(\mathbf{X})+\epsilon},
\end{equation}
where mean and standard deviation are computed across channel, time, and frequency. The feature distillation loss is
\begin{equation}
\mathcal{L}_{\mathrm{KD}}^{\mathrm{feat}}
=
\lVert
\mathrm{Norm}(\mathbf{H}^{s})
-
\mathrm{Norm}(\mathbf{H}_{\mathrm{agg}}^{t})
\rVert_2^2.
\end{equation}

\subsubsection{Progressive ramp-up and training objective}

We gradually introduce the distillation signal using
\begin{equation}
\gamma(k)
=
\min
\left(
\frac{k}{K_{\mathrm{ramp}}},
1
\right),
\end{equation}
where $k$ is the training step and $K_{\mathrm{ramp}}$ is 10\% of the total steps. Let $\mathcal{L}_{\mathrm{task}}$ denote the original speech enhancement loss, consisting of magnitude, phase, complex, time-domain, and consistency terms, following the same formulation as SEMamba~\cite{chao2024investigation}. The final objective is
\begin{equation}
\mathcal{L}_{\mathrm{total}}
=
\mathcal{L}_{\mathrm{task}}
+
\gamma(k)
\left(
\lambda_{\mathrm{out}}
\mathcal{L}_{\mathrm{KD}}^{\mathrm{out}}
+
\lambda_{\mathrm{feat}}
\mathcal{L}_{\mathrm{KD}}^{\mathrm{feat}}
\right),
\end{equation}
with
$\lambda_{\mathrm{out}}=0.5$
and
$\lambda_{\mathrm{feat}}=0.1$.

\begin{table}[t]
\centering
\caption{Results on the VCTK-DEMAND dataset for different cTF-Mamba depths and distilled students.}
\label{tab:vctk_demand}
\setlength{\tabcolsep}{4pt}
\begin{tabular}{lccccc}
\toprule
Model & PESQ & CSIG & CBAK & COVL & STOI \\
\midrule
1-layer               & 3.06 & 4.38 & 3.61 & 3.79 & 0.94 \\
2-layer              & 3.19 & 4.51 & 3.66 & 3.93 & 0.95 \\
3-layer             & 3.20 & 4.56 & 3.75 & 3.96 & 0.95 \\
4-layer            & 3.29 & 4.59 & 3.76 & 4.03 & 0.95 \\
5-layer           & 3.27 & 4.56 & 3.75 & 4.00 & 0.95 \\  
8-layer        & 3.32 & 4.64 & 3.72 & 4.08 & 0.95 \\
\hline
8-layer$\rightarrow$1-layer   & 3.18 & 4.43 & 3.68 & 3.89 & 0.95 \\
8-layer$\rightarrow$2-layer   & 3.22 & 4.55 & 3.69 & 3.97 & 0.95 \\

\bottomrule
\vspace{-4mm}
\end{tabular}
\end{table}

\begin{table*}[t]
\centering
\caption{Causal real-time speech enhancement models on VCTK-DEMAND (16 kHz).
$\dagger$ denotes models trained on larger corpora (e.g., DNS) but evaluated on VCTK-DEMAND.
$^\ast$ indicates results and latency reported from the original papers.
Latency refers to the algorithmic delay reported in the corresponding work.}
\label{tab:vb_causal}
\begin{tabular}{l c c c c c c c c c}
\toprule
Model & Year & Venue & PESQ & CSIG & CBAK & COVL & STOI & Params & Alg. Latency (ms) \\
\midrule
Noisy & -- & -- & 1.97 & 3.34 & 2.44 & 2.63 & 0.92 & -- & -- \\

PercepNet\cite{valin2020perceptually}$^\dagger$$^{\ast}$           & 2020 & Interspeech     & 2.73 & --   & --   & --   & --   & 8.00M & 40 \\
DCCRN+\cite{lv2021dccrn+}$^{\ast}$                        & 2021 & Interspeech      & 2.84 & --   & --   & --   & --   & 3.30M & -- \\
FullSubNet+\cite{chen2022fullsubnet+}$^{\ast}$                   & 2022 & ICASSP     & 2.88 & 3.86 & 3.42 & 3.57 & 0.94 & 8.67M & -- \\
DEMUCS\cite{defossez2020real}$^\dagger$$^{\ast}$              & 2020 & Interspeech     & 2.93 & 4.22 & 3.25 & 3.52 & --   & - & 40 \\
LiSenNet\cite{yan2025lisennet}$^{\ast}$                      & 2025 & ICASSP         & 3.07 & --   & --   & --   & 0.94 & 0.037M& -- \\
DeepFilterNet2\cite{schroter2022deepfilternet2}$^\dagger$$^{\ast}$      & 2022 & IWAENC    & 3.08 & 4.30 & 3.40 & 3.70 & 0.94 & 2.31M & 40 \\
FRCRN\cite{zhao2022frcrn}$^\dagger$$^{\ast}$               & 2022 & ICASSP     & 3.21 & --   & --   & --   & --   & 6.90M & 30 \\
DeepFilterNet3\cite{schroter2023deepfilternet}$^\dagger$$^{\ast}$      & 2023 & Interspeech& 3.17 & 4.34 & 3.61 & 3.77 & 0.94 & 2.13M & 40 \\
aTENNuate (base)\cite{pei2024atennuate}$^\dagger$$^{\ast}$    & 2025 & Interspeech         & 3.27 & --   & --   & --   & --   & 0.84M & 46.5 \\
\midrule
RT-SEMamba (8$\rightarrow$1-layer) & -- & Ours & 3.18 & 4.43 & 3.68 & 3.89 & 0.95 & 1.05M & 25 \\
RT-SEMamba (8$\rightarrow$2-layer) & -- & Ours & 3.22 & 4.55 & 3.69 & 3.97 & 0.95 & 1.29M & 25 \\
\textbf{RT-SEMamba (8-layer)}     & -- & Ours & \textbf{3.32} & 4.64 & 3.72 & 4.08 & 0.95 & 2.74M & 25 \\
\bottomrule
\vspace{-4mm}
\end{tabular}
\end{table*}

\begin{table}[t]
\centering
\caption{Teacher architecture search on VCTK-DEMAND. ``cTF-Trans.'' indicates replacing one cTF-Mamba block with a cTF-Transformer; otherwise all blocks are cTF-Mamba. The 8-layer all-Mamba model is used as the teacher.}
\label{tab:vctk_teacher_search}
\begin{tabular}{cc|cccc}
\toprule
Depth & cTF-Trans. & PESQ & CSIG & CBAK & COVL \\
\midrule
3 & --        & 3.20 & 4.56 & 3.75 & 3.96 \\
3 & 2nd block   & 3.19 & 4.52 & 3.68 & 3.93 \\
5 & --        & 3.27 & 4.56 & 3.75 & 4.00 \\
5 & 4th block   & \textbf{3.32} & 4.62 & \textbf{3.78} & 4.07 \\
8 & -- & \textbf{3.32} & \textbf{4.64} & 3.72 & \textbf{4.08} \\
\bottomrule
\vspace{-4mm}
\end{tabular}
\end{table}



\section{Experiments}
\label{sec:exp}

\subsection{Dataset and evaluation protocol}
\label{ssec:dataset}

We conduct experiments on the VCTK-DEMAND dataset~\cite{valentini2016investigating}, a widely
used benchmark for single-channel speech enhancement. Following the
standard configuration, the training set is created by mixing clean
utterances from 28 speakers with 10 noise types from the DEMAND
corpus at four SNR levels (0, 5, 10, and 15 dB), yielding 11{,}572
noisy--clean pairs. The test set consists of 824 utterances from two
unseen speakers mixed with five unseen noise types at four SNR levels
(2.5, 7.5, 12.5, and 17.5 dB). All signals are resampled to 16 kHz for training and evaluation, following the same setup as in~\cite{CMGAN,chao2024investigation}, including the SNR ranges.

\subsection{Quality–latency trade-off of cTF-Mamba}
\label{ssec:depth_scaling}

We first study how the number of cTF-Mamba blocks affects enhancement quality and streaming cost. Table~\ref{tab:vctk_demand} summarizes objective scores for 1–8 layers. PESQ, CSIG, and COVL generally increase with depth, while STOI saturates quickly. The largest gain comes from 1 to 2 layers (PESQ 3.06$\rightarrow$3.19); further increasing depth yields only modest improvements (2$\rightarrow$4 layers: 3.19$\rightarrow$3.29), and 5 layers no longer outperform 4 layers. The 8-layer model achieves the best overall performance (3.32 PESQ) and is therefore selected as the teacher, although the 4-, 5-, and 8-layer variants perform similarly.


Streaming cost scales almost linearly with depth: parameters grow from 1.05M to 2.74M, MACs from 20.56 to 47.35~G/s, and RTF from 0.11 to 0.29 when going from 1 to 8 layers. Thus, deeper models offer only marginal quality gains at a substantially higher real-time cost, whereas 1–2 layers are attractive for latency but noticeably weaker in quality. This motivates compressing the 8-layer teacher into very shallow students via knowledge distillation (KD).

\subsection{Knowledge distillation: 8-layer \texorpdfstring{$\rightarrow$}{→} 1-layer}
\label{ssec:kd_main}

We now evaluate KD from the 8-layer cTF-Mamba teacher to lightweight 1- and 2-layer students. The encoder, decoder, and front-end are shared; the student keeps 1 or 2 cTF-Mamba blocks (initialized from the teacher) and is trained with the same task loss plus KD losses on complex spectral outputs and intermediate features.

Table~\ref{tab:vctk_demand} also compares shallower baselines against their distilled counterparts. For the 1-layer student, KD improves PESQ from 3.06 to 3.18. The 2-layer distilled model likewise outperforms the naive 2-layer baseline (3.19$\rightarrow$3.22 PESQ), bringing its performance close to that of the 3-layer model.

To quantify transfer from the teacher, we define for a metric GapRecovery($M$), where
\[
\mathrm{GapRecovery}(M) =
\frac{M_{\text{KD}} - M_{\text{student}}}{M_{\text{teacher}} - M_{\text{student}}}
\times 100\%.
\]

For the 1-layer student, KD recovers 46.2\%, 19.2\%, 63.6\%, and 34.5\% of the teacher–student gap in PESQ, CSIG, CBAK, and COVL, respectively. Importantly, these gains incur no additional streaming cost: the distilled 1-layer student preserves the 0.11 RTF of the naive 1-layer baseline while running about $2.64\times$ faster than the 8-layer teacher (0.29 RTF). The 2-layer KD student similarly remains in a low-RTF regime ($\approx0.13$).

\begin{figure}[t]
    \centering
    \centerline{\includegraphics[width=8.6cm]{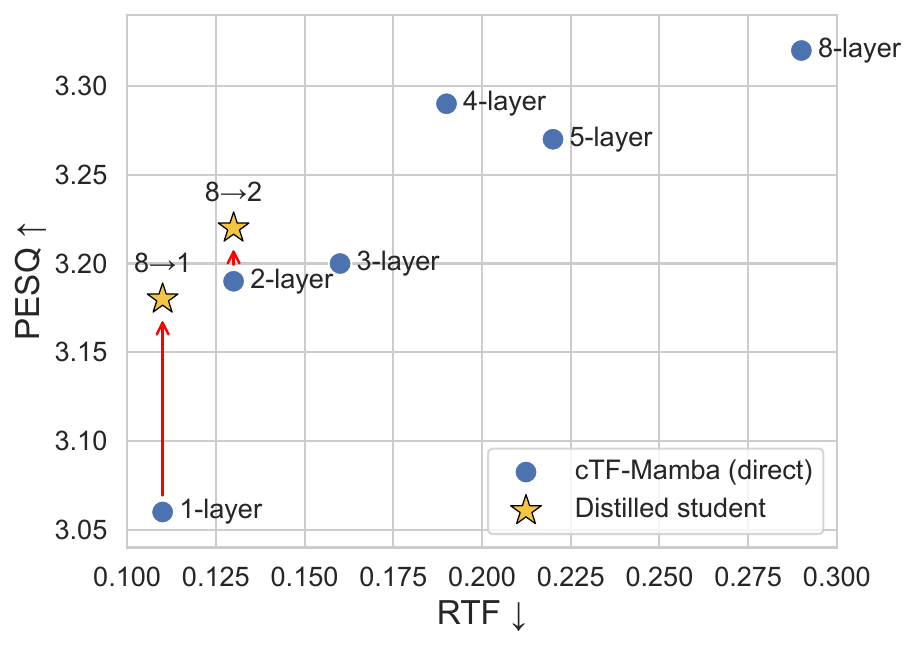}}
    \caption{Distillation improves quality without increasing latency.}
    \label{fig:teasor}
    \vspace{-4mm}
\end{figure}

Fig.~\ref{fig:teasor} visualizes this quality–efficiency trade-off. While increasing model depth improves PESQ at the cost of higher RTF, distillation shifts the operating point upward in quality without increasing runtime. In particular, the distilled 1-layer model achieves a substantial quality gain over the direct 1-layer model while maintaining identical RTF, demonstrating that KD effectively improves the Pareto frontier of streaming SE models.

\subsection{Ablation studies}
\label{ssec:ablation}

We analyze two aspects of the proposed design: (i) alternative teacher
architectures that mix Mamba and Transformer blocks, and (ii) the
contribution of individual KD components.

\subsubsection{Hybrid Mamba--Transformer teachers}
\label{sssec:jamba}


We also further explore hybrid Mamba--Transformer teachers. Jamba~\cite{lieber2024jamba} is a large language model that interleaves Mamba and Transformer blocks to combine long-context efficiency with expressive attention. Following this design, we replace one cTF-Mamba block with a cTF-Transformer at a chosen depth. The cTF-Transformer uses a causal Time-Transformer with future-masked self-attention and a 0.5~s key--value (KV) cache at inference, thus satisfying the same streaming constraints but requiring explicit cache management. As shown in Table~\ref{tab:vctk_teacher_search}, for 3-layer teachers, swapping the 2nd block yields virtually no gain (3.20 vs.\ 3.19 PESQ) and slightly degrades CBAK and COVL. For 5-layer teachers, replacing the 4th block clearly helps: PESQ improves from 3.27 to 3.32, bringing the 5-layer hybrid very close to the 8-layer all-Mamba model (3.32 PESQ).

Thus, a single Transformer block can substantially boost a mid-depth model, effectively letting a 5-layer hybrid match the 8-layer teacher, but at the cost of maintaining a KV cache in addition to the recurrent Mamba state. To avoid this extra complexity in our streaming implementation, we adopt the 8-layer all-Mamba model as the teacher in this study.

\begin{figure}[t]
    \centering
    \centerline{\includegraphics[width=9.1cm]{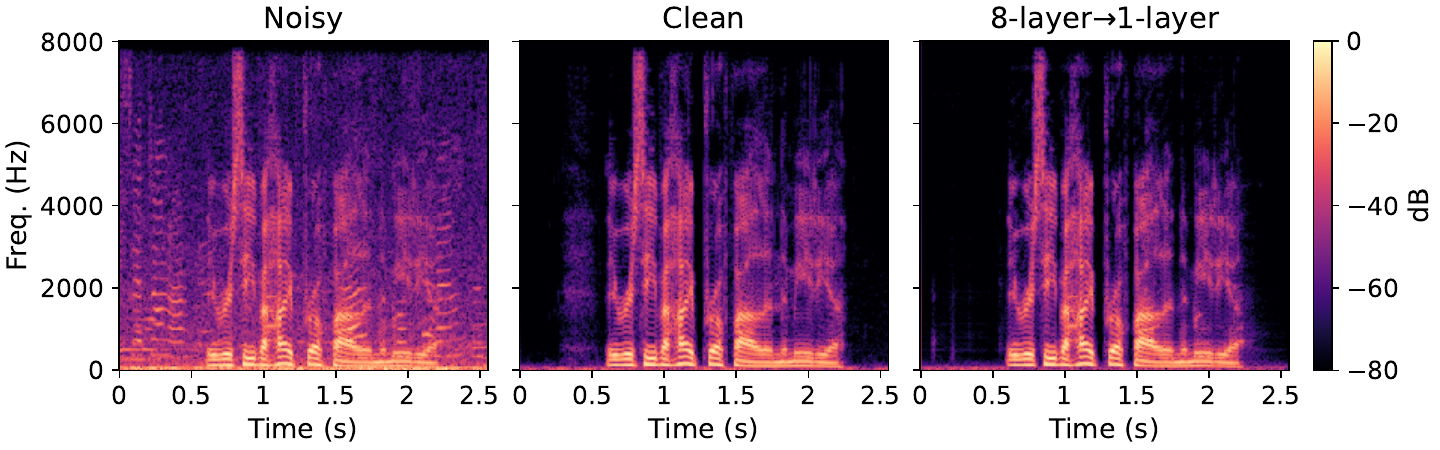}}
    \caption{Spectrogram visualization of noisy, clean, and enhanced speech produced by the distilled RT-SEMamba (8$\rightarrow$1).}
    \label{fig:visual}
    \vspace{-6mm}
\end{figure}

\subsection{Comparison with prior causal and real-time models}
\label{ssec:causal_compare}

Finally, we compare RT-SEMamba with prior causal or real-time speech
enhancement systems on VCTK-DEMAND, summarized in Table~\ref{tab:vb_causal}. Experiment result shows that RT-SEMamba achieves competitive performance under a strict latency budget. The 1-layer KD student (8$\rightarrow$1) reaches 3.18 PESQ with 1.05M parameters and a fixed 25~ms algorithmic latency. The 2-layer KD student (8$\rightarrow$2) further improves to 3.22 PESQ , while the 8-layer model attains 3.32 PESQ at 2.74M parameters, all with the same 25~ms delay. A qualitative spectrogram comparison is provided in Fig.~\ref{fig:visual}.

Compared to recent low-latency systems trained on larger corpora, the
proposed RT-SEMamba models offer competitive PESQ and COVL while using a
compact architecture and a strictly bounded 25~ms algorithmic latency.
This makes the 8$\rightarrow$1 and 8$\rightarrow$2 KD variants attractive for practical real-time enhancement under tight latency and data constraints.

\section{Conclusion}
\label{sec:conclusion}
We presented RT-SEMamba, a streaming speech enhancement architecture that uses causal time--frequency Mamba blocks to sidestep the growing memory footprint of Transformer key–value caches. To meet real-time and edge-computing constraints, we applied a progressive KD scheme to compress an 8-layer causal teacher into a lightweight 1-layer student. Experiments on VCTK-DEMAND show that the distilled model recovers much of the teacher–student performance gap and achieves competitive enhancement quality with a 25\,ms algorithmic delay and an RTF of 0.11, indicating that distilled state-space models are well suited for real-time speech applications. The code will be publicly released at \emph{github: https://github.com/RoyChao19477/RT-SEMamba}.

\section{Generative AI Use Disclosure}
Generative AI was used only for editing and polishing this manuscript.



\bibliographystyle{IEEEtran}
\bibliography{mybib}

@inproceedings{tan2018convolutional,
  title={A convolutional recurrent neural network for real-time speech enhancement.},
  author={Tan, K. and Wang, D.},
  booktitle={Proc. Interspeech},
  year={2018}
}

@article{li2020speech,
  title={Speech enhancement using progressive learning-based convolutional recurrent neural network},
  author={Li, A. and Yuan, M. and Zheng, C. and Li, X.},
  journal={Applied Acoustics},
  volume={166},
  pages={107347},
  year={2020},
  publisher={Elsevier}
}

@article{wang2018supervised,
  title={Supervised speech separation based on deep learning: An overview},
  author={Wang, D. and Chen, J.},
  journal={{IEEE/ACM Transactions on Audio, Speech, and Language Processing}},
  volume={26},
  number={10},
  pages={1702--1726},
  year={2018},
  publisher={IEEE}
}

@inproceedings{valentini2016investigating,
  title={{Investigating RNN-based speech enhancement methods for noise-robust text-to-speech.}},
  author={Valentini-Botinhao, C. and Wang, X. and Takaki, S. and Yamagishi, J.},
  booktitle={Proc. SSW},
  year={2016}
}

@article{gu2023mamba,
  title={Mamba: Linear-time sequence modeling with selective state spaces},
  author={Gu, A. and Dao, T.},
  journal={arXiv preprint arXiv:2312.00752},
  year={2023}
}

@inproceedings{fu2021metricgan+,
  title={{MetricGAN+}: An improved version of metricgan for speech enhancement},
  author={Fu, S.-W. and Yu, C. and Hsieh, T.-A. and Plantinga, P. and Ravanelli, M. and Lu, X. and Tsao, Y.},
  booktitle={Proc. Interspeech},
  year={2020}
}

@article{CMGAN,
  title={{CMGAN}: Conformer-based metric-GAN for monaural speech enhancement},
  author={Abdulatif, S. and Cao, R. and Yang, B.},
  journal={IEEE/ACM Transactions on Audio, Speech, and Language Processing},
  year={2024},
  volume={32},
  pages={2477--2493},
  publisher={IEEE}
}

@inproceedings{Trans,
  title={Speech enhancement using self-adaptation and multi-head self-attention},
  author={Koizumi, Y. and Yatabe, K. and Delcroix, M. and Masuyama, Y. and Takeuchi, D.},
  booktitle={Proc. ICASSP},
  year={2020}
}

@INPROCEEDINGS{LSTM, 
author={Z. Chen and S. Watanabe and H. Erdogan and J. R. Hershey}, 
title={Speech enhancement and recognition using multi-task learning of long short-term memory recurrent neural networks}, 
booktitle={Proc. Interspeech}, 
year={2015}, 
}

@article{FCN,
  title={End-to-end waveform utterance enhancement for direct evaluation metrics optimization by fully convolutional neural networks},
  author={Fu, S.-W. and Wang, T.-W. and Tsao, Y. and Lu, X. and Kawai, H.},
  journal={IEEE/ACM Transactions on Audio, Speech and Language Processing},
  volume={26},
  number={9},
  pages={1570--1584},
  year={2018}
}

@book{SE,
author = {Loizou, P. C.},
title = {{Speech Enhancement}: Theory and Practice},
year = {2013},
isbn = {1466504218},
publisher = {CRC Press},
address = {USA},
edition = {2nd}
}

@article{HA, 
title={Deep learning reinvents the hearing aid}, 
author={D. Wang}, 
journal={IEEE Spectrum}, 
year={2017}, 
volume={54}, 
number={3}, 
pages={32--37},
publisher={IEEE}
}

@article{gonzalez2024investigating,
  title={Investigating the design space of diffusion models for speech enhancement},
  author={Gonzalez, Philippe and Tan, Zheng-Hua and {\O}stergaard, Jan and Jensen, Jesper and Alstr{\o}m, Tommy Sonne and May, Tobias},
  journal={IEEE/ACM Transactions on Audio, Speech, and Language Processing},
  volume={32},
  pages={4486--4500},
  year={2024},
  publisher={IEEE}
}

@ARTICLE{Richter2023,
  author={Richter, J. and Welker, S. and Lemercier, J.-M. and Lay, B. and Gerkmann, T.},
  journal={IEEE/ACM Transactions on Audio, Speech, and Language Processing}, 
  title={Speech Enhancement and Dereverberation With Diffusion-Based Generative Models}, 
  year={2023},
  volume={31},
  pages={2351--2364}
}

@inproceedings{welker22_interspeech,
  author={S. Welker and J. Richter and T. Gerkmann},
  title={{Speech Enhancement with Score-Based Generative Models in the Complex STFT Domain}},
  booktitle={Proc. Interspeech},
  year={2022}
}

@INPROCEEDINGS{diffSE2022,
  title={Conditional Diffusion Probabilistic Model for Speech Enhancement}, 
  author={Lu, Yen-Ju and Wang, Zhong-Qiu and Watanabe, Shinji and Richard, Alexander and Yu, Cheng and Tsao, Yu},
  booktitle={Proc. ICASSP}, 
  year={2022}
}

@inproceedings{fu2019metricgan,
  title={{MetricGAN}: Generative adversarial networks based black-box metric scores optimization for speech enhancement},
  author={Fu, S.-W. and Liao, C.-F. and Tsao, Y. and Lin, S.-D.},
  booktitle={Proc. ICML},
  year={2019}
}

@inproceedings{mp_senet,
  author={Y.-X. Lu and Y. Ai and Z.-H. Ling},
  title={{MP-SENet: A Speech Enhancement Model with Parallel Denoising of Magnitude and Phase Spectra}},
  booktitle={Proc. Interspeech},
  year={2023}
}

@article{li2024spmamba,
  title={{SPMamba}: State-space model is all you need in speech separation},
  author={Li, K. and Chen, G.},
  journal={arXiv preprint arXiv:2404.02063},
  year={2024}
}

@inproceedings{chao2024investigation,
  title={An investigation of incorporating mamba for speech enhancement},
  author={Chao, Rong and Cheng, Wen-Huang and La Quatra, Moreno and Siniscalchi, Sabato Marco and Yang, Chao-Han Huck and Fu, Szu-Wei and Tsao, Yu},
  booktitle={Proc. IEEE SLT},
  year={2024}
}

@inproceedings{streamingSE,
  title={Causal speech enhancement with predicting semantics based on quantized self-supervised learning features},
  author={Tsunoo, Emiru and Saito, Yuki and Nakata, Wataru and Saruwatari, Hiroshi},
  booktitle={Proc. ICASSP},
  year={2025},
}

@inproceedings{pei2024atennuate,
  title={Optimized real-time speech enhancement with deep SSMs on raw audio},
  author={Pei, Yan Ru and Shrivastava, Ritik and Sidharth, FNU},
  booktitle={Proc. Interspeech},
  year={2025}
}

@inproceedings{schroter2023deepfilternet,
  title={{DeepFilterNet}: Perceptually motivated real-time speech enhancement},
  author={Schr{\"o}ter, Hendrik and Rosenkranz, Tobias and Escalante-B, Alberto N and Maier, Andreas},
  booktitle={Proc. Interspeech},
  year={2023}
}

@inproceedings{schroter2022deepfilternet2,
  title={{DeepFilterNet2}: Towards real-time speech enhancement on embedded devices for full-band audio},
  author={Schr{\"o}ter, Hendrik and Maier, A and Escalante-B, Alberto N and Rosenkranz, Tobias},
  booktitle={Proc. IEEE IWAENC},
  year={2022},
}

@inproceedings{zhao2022frcrn,
  title={{FRCRN}: Boosting feature representation using frequency recurrence for monaural speech enhancement},
  author={Zhao, Shengkui and Ma, Bin and Watcharasupat, Karn N and Gan, Woon-Seng},
  booktitle={Proc. ICASSP},
  year={2022},
}

@inproceedings{yan2025lisennet,
  title={{LiSenNet}: Lightweight sub-band and dual-path modeling for real-time speech enhancement},
  author={Yan, Haoyin and Zhang, Jie and Fan, Cunhang and Zhou, Yeping and Liu, Peiqi},
  booktitle={Proc. ICASSP},
  year={2025},
}

@article{defossez2020real,
  title={Real time speech enhancement in the waveform domain},
  author={Defossez, Alexandre and Synnaeve, Gabriel and Adi, Yossi},
  journal={Proc. Interspeech},
  year={2020}
}

@article{RealtimeSE,
  title={Real time speech enhancement in the waveform domain},
  author={Defossez, Alexandre and Synnaeve, Gabriel and Adi, Yossi},
  journal={Proc. Interspeech},
  year={2020}
}

@inproceedings{lv2021dccrn+,
  title={{DCCRN+}: Channel-wise subband dccrn with snr estimation for speech enhancement},
  author={Lv, Shubo and Hu, Yanxin and Zhang, Shimin and Xie, Lei},
  booktitle={Proc. Interspeech},
  year={2021}
}

@inproceedings{schroter2022deepfilternet,
  title={{DeepFilterNet}: A low complexity speech enhancement framework for full-band audio based on deep filtering},
  author={Schroter, Hendrik and Escalante-B, Alberto N and Rosenkranz, Tobias and Maier, Andreas},
  booktitle={Proc. ICASSP},
  year={2022},
}

@inproceedings{valin2020perceptually,
  title={A perceptually-motivated approach for low-complexity, real-time enhancement of fullband speech},
  author={Valin, Jean-Marc and Isik, Umut and Phansalkar, Neerad and Giri, Ritwik and Helwani, Karim and Krishnaswamy, Arvindh},
  booktitle={Proc. Interspeech},
  year={2020}
}

@article{lieber2024jamba,
  title={Jamba: A hybrid transformer-mamba language model},
  author={Lieber, Opher and Lenz, Barak and Bata, Hofit and Cohen, Gal and Osin, Jhonathan and Dalmedigos, Itay and Safahi, Erez and Meirom, Shaked and Belinkov, Yonatan and Shalev-Shwartz, Shai and others},
  journal={arXiv preprint arXiv:2403.19887},
  year={2024}
}

@inproceedings{wang2025mamba,
  title={{Mamba-SEUNet}: Mamba UNet for monaural speech enhancement},
  author={Wang, Junyu and Lin, Zizhen and Wang, Tianrui and Ge, Meng and Wang, Longbiao and Dang, Jianwu},
  booktitle={Proc. ICASSP},
  year={2025},
}

@article{ahmed2025neuroamp,
  title={{NeuroAMP}: A novel end-to-end general purpose deep neural amplifier for personalized hearing aids},
  author={Ahmed, Shafique and Zezario, Ryandhimas E and Yuan, Hui-Guan and Hussain, Amir and Wang, Hsin-Min and Chung, Wei-Ho and Tsao, Yu},
  journal={IEEE Transactions on Artificial Intelligence},
  year={2025},
  publisher={IEEE}
}

@inproceedings{li2024diffusion,
  title={Diffusion-based generative modeling with discriminative guidance for streamable speech enhancement},
  author={Li, Chenda and Cornell, Samuele and Watanabe, Shinji and Qian, Yanmin},
  booktitle={Proc. IEEE SLT},
  year={2024},
}

@article{scheibler2024universal,
  title={Universal score-based speech enhancement with high content preservation},
  author={Scheibler, Robin and Fujita, Yusuke and Shirahata, Yuma and Komatsu, Tatsuya},
  journal={Proc. Interspeech},
  year={2024}
}

@inproceedings{yang2022audio,
  title={Audio-visual speech codecs: Rethinking audio-visual speech enhancement by re-synthesis},
  author={Yang, Karren and Markovi{\'c}, Dejan and Krenn, Steven and Agrawal, Vasu and Richard, Alexander},
  booktitle={Proc. IEEE/CVF CVPR},
  year={2022}
}

@inproceedings{hsu2022learning,
  title={Learning-based personal speech enhancement for teleconferencing by exploiting spatial-spectral features},
  author={Hsu, Yicheng and Lee, Yonghan and Bai, Mingsian R},
  booktitle={Proc. ICASSP},
  year={2022},
}

@article{kolbaek2020loss,
  title={On loss functions for supervised monaural time-domain speech enhancement},
  author={Kolb{\ae}k, Morten and Tan, Zheng-Hua and Jensen, S{\o}ren Holdt and Jensen, Jesper},
  journal={IEEE/ACM Transactions on Audio, Speech, and Language Processing},
  volume={28},
  pages={825--838},
  year={2020},
  publisher={IEEE}
}

@article{o2024speech,
  title={Speech enhancement—A review of modern methods},
  author={O'Shaughnessy, Douglas},
  journal={IEEE Transactions on Human-Machine Systems},
  volume={54},
  number={1},
  pages={110--120},
  year={2024},
  publisher={IEEE}
}

@article{lai2016deep,
  title={A deep denoising autoencoder approach to improving the intelligibility of vocoded speech in cochlear implant simulation},
  author={Lai, Ying-Hui and Chen, Fei and Wang, Syu-Siang and Lu, Xugang and Tsao, Yu and Lee, Chin-Hui},
  journal={IEEE Transactions on Biomedical Engineering},
  volume={64},
  number={7},
  pages={1568--1578},
  year={2016},
  publisher={IEEE}
}

@inproceedings{yin2020phasen,
  title={Phasen: A phase-and-harmonics-aware speech enhancement network},
  author={Yin, Dacheng and Luo, Chong and Xiong, Zhiwei and Zeng, Wenjun},
  booktitle={Proc. AAAI},
  year={2020}
}

@article{hu2020dccrn,
  title={{DCCRN}: Deep complex convolution recurrent network for phase-aware speech enhancement},
  author={Hu, Yanxin and Liu, Yun and Lv, Shubo and Xing, Mengtao and Zhang, Shimin and Fu, Yihui and Wu, Jian and Zhang, Bihong and Xie, Lei},
  journal={Proc. Interspeech},
  year={2020}
}

@inproceedings{chen2022fullsubnet+,
  title={Fullsubnet+: Channel attention fullsubnet with complex spectrograms for speech enhancement},
  author={Chen, Jun and Wang, Zilin and Tuo, Deyi and Wu, Zhiyong and Kang, Shiyin and Meng, Helen},
  booktitle={Proc. ICASSP},
  year={2022},
}

@inproceedings{hao2021fullsubnet,
  title={Fullsubnet: A full-band and sub-band fusion model for real-time single-channel speech enhancement},
  author={Hao, Xiang and Su, Xiangdong and Horaud, Radu and Li, Xiaofei},
  booktitle={Proc. ICASSP},
  year={2021},
}

@article{hinton2015distilling,
  title={Distilling the knowledge in a neural network},
  author={Hinton, Geoffrey and Vinyals, Oriol and Dean, Jeff},
  journal={arXiv preprint arXiv:1503.02531},
  year={2015}
}

@article{chao2025universal,
  title={Universal speech enhancement with regression and generative mamba},
  author={Chao, Rong and Nasretdinov, Rauf and Wang, Yu-Chiang Frank and Juki{\'c}, Ante and Fu, Szu-Wei and Tsao, Yu},
  journal={Proc. Interspeech},
  year={2025}
}

@inproceedings{avenstrup2025sepmamba,
  title={SepMamba: State-space models for speaker separation using Mamba},
  author={Avenstrup, Thor H{\o}jhus and Elek, Boldizs{\'a}r and M{\'a}di, Istv{\'a}n L{\'a}szl{\'o} and Schin, Andr{\'a}s Bence and M{\o}rup, Morten and Jensen, Bj{\o}rn Sand and Olsen, Kenny},
  journal={Proc. ICASSP},
  year={2025},
}

@article{kim2025mamba,
  title={Mamba-based Hybrid Model for Speech Enhancement},
  author={Kim, Se-Ha and Kim, Tae-Gyeong and Chun, Chang-Jae},
  journal={Proc. Interspeech},
  year={2025}
}

\end{document}